\documentstyle[amstex,amssymb,preprint,aps]{revtex}

\begin{document}
\title{Anomalous oxygen isotope effect on the in-plane FIR conductivity of
detwinned YBa$_{2}$Cu$_{3}^{16,18}$O$_{6.9}$.}
\author{C. Bernhard$^{1)}$, T. Holden$^{1)}$, A.V. Boris$^{1)}$, N.N.
Kovaleva$^{1)}$, A.V. Pimenov$^{1)}$, J. Humlicek$^{2)}$, C. Ulrich$^{1)}$,
C.T. Lin$^{1)}$, and J.L. Tallon$^{3)}$}
\address{1) Max-Planck-Institut for Solid State Research, Heisenbergstrasse\\
1, D-70569 Stuttgart, Germany.\\
2) Masaryk Universiy, Brno, Czech Republic\\
3) Industrial Research Laboratory, Lower Hutt, New Zealand}
\maketitle

\begin{abstract}
We observe an anomalous oxygen isotope effect on the a-axis component of the
far-infrared electronic response of detwinned YBa$_{2}$Cu$_{3}^{16,18}$O$%
_{6.9}$. For $^{18}$O a pronounced low-energy electronic mode (LEM) appears
around 240 cm$^{-1}$. This a-axis LEM exhibits a clear aging effect, after
one year it is shifted to 190 cm$^{-1}$. For $^{16}$O we cannot resolve a
corresponding a-axis LEM above 120 cm$^{-1}$. We interpret the LEM in terms
of a collective electronic mode that is pinned by `isotopic defects', i.e.
by the residual $^{16}$O in the matrix of $^{18}$O.
\end{abstract}


There is an ongoing debate on whether electron-phonon coupling is at the
heart of (at least some of) the unusual electronic properties of the cuprate
high-T$_{c}$ superconductors (HTSC). The fact that the superconducting (SC)
transition temperature, T$_{c}$, exhibits only a very small isotope effect,
especially in optimally doped samples with the highest T$_{c}$ values \cite%
{Franck1} has led many researchers to neglect electron-phonon coupling.
Nevertheless, some recent experiments favor a strong electron-phonon
interaction. For example, it has been argued that a kink in the electronic
dispersion, as measured by angle-resolved photo-emission (ARPES), arises due
to strong electron-phonon interaction \cite{Shen1}. Most notably, it was
reported that some electronic properties exhibit a sizeable oxygen isotope
effect (OIE). In particular, a very large OIE has been reported for the
onset temperature, T$^{\ast }$, of the pseudogap phenomenon in underdoped
HTSC \cite{Temprano1,Lanzara1} which manifests itself as a suppression of
the low-energy spin- and charge excitations already in the normal state
(NS). However, the experimental results are still controversial, ranging
from only a marginal OIE as seen by nuclear magnetic resonance (NMR) \cite%
{Williams1,Raffa1} to a giant one as obtained by inelastic neutron
scattering (INS) utilizing crystal field excitations \cite{Temprano1} and by
x-ray absorption spectroscopy (XAS) \cite{Lanzara1}. The significance and
the implications of these conflicting results are subject to an ongoing
controversial discussion.

The present experiments were motivated by the idea that the huge variation
in the OIE may be due to the different time-scales of these techniques,
i.e., 10$^{-5}$ to 10$^{-8}$ s in NMR as compared with 10$^{-12}$-10$^{-15}$
in INS\ and XAS. In that case one might expect that the far-infrared (FIR)
electronic response exhibits a sizeable OIE since it involves a time scale
of 10$^{-12}$-10$^{-14}$ s (1 THz$\triangleq 33\;cm^{-1}$). So far only few
reports exist on the OIE of the IR-electronic response that were obtained on
polycrystalline samples or on strongly underdoped twinned crystals \cite%
{Taliani1,Timusk1}. These studied do not give a conclusive answer as to the
OIE\ of the FIR electronic response. Accordingly, we performed FIR
ellipsometric measurements on detwinned and optimally doped YBa$_{2}$Cu$%
_{3}^{16,18}$O$_{6.9}$ crystals with either $^{16}$O or $^{18}$O. To our
surprise we find indeed that the a-axis component of the electronic response
exhibits sizeable changes upon isotopic substitution. For $^{18}$O a
pronounced low-energy electronic mode (LEM) occurs around 240 cm$^{-1}$ at
10 K. Moreover, this a-axis LEM exhibits a surprising aging effect. After
one year it has shifted to a lower frequency of $\sim $190 cm$^{-1}$. In the 
$^{16}$O sample a corresponding LEM is not seen above 120 cm$^{-1}$. Either
it occurs at lower frequencty or it is fully absent. We interpret our data
in terms of a collective electronic mode that becomes pinned by isotopic
defects. The isotopic defects most likely arise due to the few percent of
residual $^{16}$O in the $^{18}$O substituted crystal.

Pairs of flux-grown Y-123 crystals were selected from the same growth batch
and annealed in isotopically enriched oxygen under identical conditions to
obtain optimally doped crystals (final annealing step for 7 days at 490 $%
{{}^\circ}%
$C in O$_{2}$) \cite{Tallon1}. Two pairs were detwinned using a home-built
apparatus that allows application of mechanical pressure while the crystal
is heated in an enclosed atmosphere of either $^{16}$O$_{2}$ or $^{18}$O$%
_{2} $ gas. Complete detwinning was confirmed with a polarization microscope
and by x-ray diffraction. The sample surfaces were polished to optical grade
using diamond paste. The high quality of these surfaces was confirmed by
Raman measurements. The crystals exhibit sharp SC transitions at T$_{c}$=92.7%
$\pm 0.5$ K for $^{16}$O and 92.5$\pm 0.4$ K for $^{18}$O as determined by
dc SQUID magnetization.

The ellipsometric measurements were performed at the U4IR beamline at NSLS
in Brookhaven, USA and at ANKA in Karlsruhe, Germany \cite{Golnik1}.
Ellipsometry has the advantage that it measures directly the complex
dielectric function without a need for reference measurements or
Kramers-Kronig analysis. Details of the techniques are given in \cite%
{Golnik1,Humlicek1}.

Figure 1 displays the spectra for the a-axis component (perpendicular to the
CuO chains) of the freshly prepared $^{18}$O-substituted crystal. Shown are
the real parts of (a) the conductivity, $\sigma _{1a}$=1/4$\pi \cdot \nu
\cdot \varepsilon _{2a},$ and (b) the dielectric function, $\varepsilon
_{1a} $. The most remarkable feature is a pronounced mode around 240 cm$%
^{-1} $ (at 10 K) whose large spectral weight (SW) immediately suggests that
it is of electronic origin. The parameters of this low-energy electronic
mode (LEM) have been obtained with the function:

$\varepsilon \left( \nu \right) =\varepsilon _{\infty }+\frac{S_{LEM}\cdot
\nu ^{2}}{(\nu ^{2}-\nu _{LEM}^{2})+i\cdot \nu \cdot \Gamma _{LEM}}-\frac{%
\nu _{D}^{2}}{\nu ^{2}+i\cdot \nu \cdot \Gamma _{D}}+%
\mathrel{\mathop{\sum }\limits_{i=1,2}}%
\frac{S_{i}\cdot \nu ^{2}}{(\nu ^{2}-\nu _{i}^{2})+i\cdot \nu \cdot \Gamma
_{i}}$. (1)

The LEM is described by a Lorentz oscillator of strength, S$_{LEM}$, center
frequency, $\nu _{LEM},$ and width, $\Gamma _{LEM}$. The remaining part of
the electronic response is accounted for by a Drude-term with plasma
frequency, $\nu _{D}=\sqrt{n\cdot e^{2}/\pi \cdot m^{\ast }},$ carrier
density, n, and effective mass, m*, and by two broad Lorentzians with $%
\Gamma _{1,2}$=const=1000 cm$^{-1}$ to reproduce the mid-infrared (MIR)
band. For simplicity we did not include the IR active phonons. We obtained S$%
_{LEM}$=692(8), 596(8) and 438(3), $\nu _{LEM}=$238.7(0.7), 245(1.5) and
285(2.4) cm$^{-1}$ and $\Gamma _{LEM}$=134(7), 155(10), and 197(6) cm$^{-1}$
at 10, 100 and 300 K respectively. The parameters of the Drude-term are $\nu
_{D}$=10500(700) cm$^{-1}$ (nearly T-independent) and $\Gamma _{D}=$35, 42
and 98 cm$^{-1}$ at T=10, 100 and 300 K. The fit functions are shown in the
inset of Fig. 1 by the dashed lines. The SW of the LEM amounts to almost 10
\% of the total SW of the charge carriers as derived from the unscreened
plasma frequency of $\sim 16000$ cm$^{-1}$ (this includes the MIR-band).\
The a-axis LEM therefore represents a significant fraction of the SW of the
charge carriers irrespective of the details of the modelling of the
electronic background. The LEM is most pronounced at 10 K but it persists
even at room temperature. Notably, its center frequency, $\nu _{LEM}$,
increases as the temperature is raised.

Figure 2a compares the a-axis spectra at 10 K for the $^{18}$O (thick solid
line) and the $^{16}$O (thin solid line) samples and highlights the large
OIE of the electronic response. Ellipsometric data for a similar $^{16}$O
crystal were previously reported \cite{Bernhard1}. The LEM\ that dominates
the a-axis response for $^{18}$O is probably shifted to much lower frequency
for $^{16}$O. Though we cannot exclude the possibility that the LEM is
absent for $^{16}$O, we believe that the strong upturn of $\sigma _{1a}$($%
\nu $) below 250 cm$^{-1}$ is related to a LEM centered around 50-100 cm$%
^{-1}$ rather than to a Drude peak at zero frequency. A very narrow Drude
peak with $\Gamma _{D}\sim 5-30cm^{-1}$\ is commonly observed in the
microwave spectral range, even at T%
\mbox{$<$}%
\mbox{$<$}%
T$_{c}$ \cite{Hardy1}, but it cannot account for the broader upturn of $%
\sigma _{1a}$($\nu $) in the FIR.

The spectra in Figs. 1 and 2 also contain some narrow modes due to IR-active
phonons \cite{Bernhard1}. The apparent red shift of the phonons for $^{18}$O
confirms the high degree of isotopic substitution ($\sim $95 \%). A detailed
account of the phonon parameters will be given elsewhere. We only mention
the interesting trend concerning the anomalously large SW of the phonons in
the vicinity of the LEM. For $^{16}$O we have previously shown that the
modes at 190 and 230 cm$^{-1},$ that are located near the upturn of $\sigma
_{1a}$($\nu $,10 K), acquire an anomalously large SW \cite{Bernhard1}. For $%
^{18}$O, where the LEM dominates the electronic response up to 400 cm$^{-1},$
not only the modes at 180 and 220 cm$^{-1}$ but also the ones at 275 and 345
cm$^{-1}$ exhibit an enhanced SW. This signals that the IR-active phonons
interact strongly with the LEM but not so much with the remaining part of
the electronic background.

After a period of about one year we repeated the ellipsometry measurements
on the $^{18}$O crystal. Surprisingly, we found that the a-axis LEM had
undergone significant changes. As is shown in Fig. 2(b), the center of the
LEM has shifted significantly upon aging towards lower frequency. Using eq.
(1) we obtain S$_{LEM}$=920(40), $\nu _{LEM}$=185(5) cm$^{-1}$ and $\Gamma
_{LEM}$=120(10) cm$^{-1}$ at 10 K. The Drude-term remains almost unchanged.
We repolished and remeasured the crystal and ensured that these changes of
the LEM cannot be an artifact due to a poor surface quality. We also
reinvestigated the aged $^{16}$O crystal (and several others) but did not
observe any significant changes of the FIR response. Finally, we repeated
the Raman-, x-ray, and SQUID\ magnetisation measurements on the aged $^{18}$%
O crystal and did not observe any noticeable changes with respect to the
freshly prepared state. In particular, the aged $^{18}$O crystal was still
fully detwinned and (except for the isotopic shift of the phonon modes)
similar Raman spectra were obtained for the $^{16}$O and $^{18}$O crystals.

Finally, Figure 2(c) compares the b-axis spectra of $\sigma _{1b}$($\nu ,10K$%
) parallel to the 1-d CuO chains for $^{16}$O and $^{18}$O. In contrast to
the a-axis LEM, the b-axis electronic response exhibits no anomalous OIE. A
similar b-axis LEM occurs for $^{16}$O and $^{18}$O, only the narrow peaks
due to the IR-active phonons are red shifted for $^{18}$O as expected due to
the mass change. Using eq. (1) we obtain similar LEM parameters of S$_{LEM}$%
=900-1000, $\nu _{LEM}$=253-258 cm$^{-1}$ and $\Gamma _{LEM}$=200-240 cm$%
^{-1}$ at 10 K.

The combined set of data, in particular, the observed aging effects suggest
that we are not dealing with a true `isotope effect'. Instead it appears
that defects related to the isotopic replacement are responsible for these
marked changes of the FIR electronic response. We analyzed the $^{18}$O gas
with a mass spectrometer and investigated a pair of $^{16}$O and $^{18}$O
crystals by combined differential thermo-gravimetry and mass-spectrometry
analysis. We obtained no evidence for extrinsic impurities that are
incorporated during the oxygen exchange for $^{18}$O but not for $^{16}$O.
We also performed a back exchange experiment for twinned Y-123 crystals and
found the isotopic effect to be reversible.

Therefore, we consider it most likely that the remaining few percent of $%
^{16}$O in the $^{18}$O substituted sample act as `isotopic defects' which
interact strongly with the electronic excitations. One possibility is that
they scatter the free charge carriers within the CuO$_{2}$ planes and thus
lead to localization. Such a scenario was previously used to explain a
similar LEM in the a-axis response of Zn-substituted Y-124 \cite{Basov1}.
However, strong scattering of the free carriers in the presence of a d--wave
order parameter would give rise to a significant reduction of T$_{c}$ and
even more so of the SC condensate density, n$_{s}$ \cite{Bernhard2}. Both
effects are hardly present in our $^{18}$O crystal with T$_{c}$=92.5 K and $%
\lambda _{a}$ $\approx $ 1600-1700 \AA\ as deduced from the zero-frequency
limit of $\lambda _{a}(v)=$[$4\cdot \pi ^{2}\cdot \nu ^{2}\cdot
(1-\varepsilon _{1}(\nu ))]^{-1/2}$ (not shown). A measurable but very small
OIE\ on the magnetic penetration depth was obtained in a previous study \cite%
{Hofer1}.

In our opinion the most likely explanation is that the `isotopic defects',
i.e. the remaining few percent of $^{16}$O in $^{18}$O, act as pinning
centers for a collective electronic mode. A possible candidate is an
incommensurate charge density wave (CDW) that involves the flat parts of the
Fermi-suface around the so-called `hot spots' close to the X-point of the
Brillouin-zone where the nesting condition is most favorable \cite%
{Castellano1}. Corresponding LEM's frequently occur in low-dimensional
materials with a CDW ground state where the phase mode is pinned by lattice
defects \cite{Gruener}. In the strong-pinning limit the frequency of the
phase mode, $\Omega _{o}$, is determined by the potential, V$_{i}$, and
concentration, n$_{i},$ of the pinning centers and by the effective mass m*
of the sliding CDW, $\Omega _{o}\approx \sqrt{\frac{V_{i}\cdot n_{i}}{m\ast }%
}$ \cite{Gruener}. In conventional CDW\ materials $\Omega _{o}$ is located
well below the FIR range. Nevertheless, a pinned CDW in the FIR was recently
observed in the ladder-compound Sr$_{14}$Cu$_{24}$O$_{41}$ at 12 cm$^{-1}$ %
\cite{Gorshunov1} and around 100 cm$^{-1}$ in Ca doped Sr$_{6}$Ca$_{8}$Cu$%
_{24}$O$_{41}$ \cite{Osafune1}. Moreover, for the cuprates there exist
several reports of a LEM around 100 cm$^{-1}$, especially for samples with
defects \cite{Basov1,Lupi1,McGuire1,Singley1}. A related model is the stripe
(fluctuation) model which assumes that, due to frustration of magnetic
exchange, the charges tend to segregate and form quasi one-dimensional
hole-rich rivers that are separated by hole-poor stripes where the AF order
of the Cu spins is maintained \cite{Kievelson1}. The existence of a LEM in
the FIR has recently been studied theoretically \cite{Benfatto} and
experimental data on La$_{2-x}$Sr$_{x}$CuO$_{4}$ have been discussed along
these lines \cite{Lucarelli}. The LEM\ corresponds here to transverse
oscillations of the individual stripes. Pinning centers decrease the length
of the stripe segments and thereby increase $\nu _{LEM}$ \cite{Benfatto}.

Unlike defect atoms, the isotopic defects will not affect the static local
electronic properties or lattice distortions. Nevertheless, the local
difference in the vibrational energy in the vicinity of isotopic defects may
well lead to the pinning a collective electronic mode that is coupled to
lattice vibrations. Based on theoretical calculations it has even been
sugggest that an incommensurate CDW\ becomes stabilized by the lattice
vibrations in near optimal doped samples \cite{Castellano1}. Accordingly,
the higher vibrational energy in the vicinity of $^{16}$O may well give rise
to an attractive pinning potential for a CDW or for stripes. An estimate of
the pinning potential V$_{i}$ can be obtained from the red-shift of the
highest IR-active phonon mode of $\sim $ 25 cm$^{-1}\triangleq $ 3 meV. From
the purity of the $^{18}$O gas and the shift of the high energy phonon modes
in the IR and\ Raman spectra we deduce a concentration of $^{16}$O `isotopic
defects' of about 5 \%. Natural oxygen contains only 0.2 \% of $^{18}$O.
This order of magnitude difference in the concentration of `isotopic
defects' thus can account for a three-fold decrease of $\Omega _{o}$ from $%
\sim $240 cm$^{-1}$ for $^{18}$O to $\sim $80 cm$^{-1}$ for $^{16}$O in
agreement with our data. The aging effect of the $^{18}$O crystal can be
understood due to a rearrangement (e.g. a clustering) of the $^{16}$O
defects within the CuO$_{2}$ planes. Moreover, the concentration of $^{16}$O
defects within the CuO$_{2}$ planes may have decreased with time due to a
gradual exchange of oxygen between the CuO$_{2}$ planes and the CuO chains
whose initial $^{18}$O content likely was somewhat higher.

Based on our present data we cannot easily distinguish between a CDW and the
stripe scenario. An important aspect is the anisotropy of the LEM which
should be very pronounced for the stripe scenario. The situation is however
complicated for Y-123 which in addition contains the one-dimensional CuO
chains that are conductive along the b-axis direction. It was previously
noted that the b-axis LEM in Y-123 most likely originates from a CDW within
the CuO chains that is pinned by structural defects like oxygen vacancies %
\cite{Bernhard1}. The presence of additional strong pinning centers within
the CuO chains can indeed explain why the b-axis LEM\ exhibits no
significant OIE. Similar experiments of the OIE of the FIR electronic
response in tetragonal systems are in progress in order to the interesting
problem of the anisotropy of the OIE of the LEM.

In summary, our ellipsometric data provide compelling evidence for a very
unusual oxygen isotope effect on the FIR\ electronic conductivity along the
a-axis. In particular, the observed aging effect suggests that we are not
witnessing a true isotope effect but rather the result of `isotopic
defects'. We suggest that the remaining few percent of $^{16}$O within the $%
^{18}$O matrix act as pinning centers for a collective electronic mode.
Possible candidates are a CDW or stripe fluctuation as suggested in Ref. %
\cite{Kievelson1,Benfatto}. While we do not arrive at a conclusive answer
concerning the nature of the collective mode, our data provide some
important insights. First of all, they identify the energy scale that is
underlying the OIE on the electronic properties with $\nu _{LEM}\sim $%
100-250 cm$^{-1}$($\sim $12-30 meV). This resolves previous conflicting data
in terms of the energy scales that are probed by the different techniques.
Secondly, our data signal that the OIE exhibits a clear aging effect on the
time scale of a year. Finally, our new data show that a significant fraction
of the electronic spectral weight is contained in a LEM which does not
contribute to the SC condensate density. The LEM can be affected in various
ways while the impact on the SC\ state, e.g. on T$_{c}$ or n$_{s},$ remains
surprisingly small. While this suggests that the LEM may not be of much
relevance for the SC\ pairing interaction, it remains to be seen whether it
is closely related to the pseudogap phenomenon which so far has prevented us
from obtaining a unique description of the SC\ state.

\bigskip

Figure 1: A-axis component of the real part of (a) the conductivity, $\sigma
_{1a}$($\nu $), and (b) the dielectric function $\varepsilon _{1a}$($\nu $)
of YBa$_{2}$Cu$_{3}^{18}$O$_{6.9}$ at 300 K (dotted line), 100 K (thin solid
line) and 10 K (thick solid line). The insets show the low-frequency range
on an enlarged scale. The dashed lines show the fits with the function in
eq. (1).

Figure 2: Comparison of $\sigma _{1}$($\nu ,$10 K) for the a-axis component
of (a) the $^{16}$O and the $^{18}$O crystals, (b) the $^{18}$O cyrstal in
the freshly prepared state and in the aged state ($\sim $1 year), and (c) of
the b-axis component parallel to the CuO chains for $^{16}$O and the $^{18}$%
O.


\begin{references}
\bibitem{Franck1} J.R. Franck, in ''Physical Properties of High Temperature
Superconductors IV'', edited by D.M. Ginsberg (World Scientific, Singapore,
1994), p.189.

\bibitem{Shen1} Z.X. Shen et al., Philos. Mag. B 82, 1349 (2002).

\bibitem{Temprano1} D. Rubio Temprano et al., Phys. Rev. Lett. {\bf 84},
1990 (2000).

\bibitem{Lanzara1} A. Lanzara et al., J. Phys. Condens. Matter {\bf 11},
L541 (1999).

\bibitem{Williams1} G.V.M. Williams et al., Phys. Rev. Lett. {\bf 80}, 377
(1998).

\bibitem{Raffa1} F. Raffa et al., Phys. Rev. Lett. {\bf 81}, 5912 (1998).

\bibitem{Taliani1} V.N. Denisov et al. Phys. Rev. {\bf B 48}, 16714 (1993).

\bibitem{Timusk1} N.L. Wang et al., Phys. Rev. Lett. {\bf 89}, 087003 (2002).

\bibitem{Tallon1} D.J. Pringle, Phys. Rev. {\bf B 62}, 12527 (2000).

\bibitem{Golnik1} A. Golnik et al., Phys. Stat. Sol. {\bf (b) 215}, 553
(1999).

\bibitem{Humlicek1} J. Huml{\'{\i}}{\v{c}}ek et al., unpublished.

\bibitem{Bernhard1} C. Bernhard et al., Solid State Commun. {\bf 121}, 93
(2002).

\bibitem{Hardy1} A. Hosseini et al., Phys. Rev. {\bf B 60}, 1349 (1999), A.
Pimenov et al., Phys. Rev. {\bf B 59}, 4390 (1999).

\bibitem{Basov1} D.N. Basov, B. Dabrowski, and T. Timusk, Phys. Rev. Lett. 
{\bf 81}, 2132 (1998).

\bibitem{Bernhard2} C. Bernhard et al., Phys. Rev. Lett. {\bf 77}, 2304
(1996).

\bibitem{Hofer1} G.M. Zhao, et al., Nature {\bf 385}, 236 (1997); J. Hofer
et al., Phys. Rev. Lett. {\bf 84}, 4192 (2000).

\bibitem{Castellano1} C. Castellano et al., Phys. Rev. Lett. {\bf 75}, 4650
(1995), F. Becca et al., Phys. Rev B 54, 12446 (1996).

\bibitem{Gruener} G. Gr\"{u}ner, Rev. Mod. Phys. {\bf 60}, 1129 (1985).

\bibitem{Gorshunov1} B. Gorshunov et al., Phys. Rev. {\bf B 66}, 060508
(2002).

\bibitem{Osafune1} T. Osafune et al., Phys. Rev. Lett. 82, 1313 (1999).

\bibitem{Lupi1} S. Lupi et al., Phys. Rev. {\bf B 62}, 12418 (2000).

\bibitem{McGuire1} J.J. McGuire et al., Phys. Rev. {\bf B 62}, 8711 (2000).

\bibitem{Singley1} J.E. Singley et al., Phys. Rev. {\bf B 64}, 224503 (2001).

\bibitem{Kievelson1} M.I. Salkola, V.J. Emery, and S.A. Kievelson, Phys.
Rev. Lett. {\bf 77}, 155 (1996); M. Ichioka and K. Machida, J. Phys. Soc.
Jpn. {\bf 68}, 4020 (1999).

\bibitem{Benfatto} L. Benfatto and C. Morasis Smith, cond-mat/0303036.

\bibitem{Lucarelli} A. Lucarelli et al., Phys. Rev. Lett. {\bf 90}, 037002
(2003), M. Dumm et al., Phys. Rev. Lett. {\bf 88}, 147003 (2002).
\end{references}
\end{document}